\documentclass{PoS}

\usepackage{dsfont}
\usepackage{slashed}
\usepackage{booktabs}
\usepackage{graphicx}
\usepackage{float}

\newcommand{\Op}{\mathcal{O}} 
\newcommand{\eins}{\mathds{1}} 

\title{Electromagnetic form factors of the $\Delta$ baryon}

\ShortTitle{Electromagnetic form factors of the $\Delta$ baryon}

\author{C. Alexandrou\\
        Department of Physics, University of Cyprus, P.O. Box 20537, 1678 Nicosia, Cyprus\\ 
        E-mail:           \email{alexand@ucy.ac.cy}}

\author{\speaker{T. Korzec}\\
        Department of Physics, University of Cyprus, P.O. Box 20537, 1678 Nicosia, Cyprus\\
        E-mail: \email{korzec@ucy.ac.cy}}

\author{T. Leontiou\\
        Department of Physics, University of Cyprus, P.O. Box 20537, 1678 Nicosia, Cyprus\\
	E-mail: \email{t.leontiou@ucy.ac.cy}}

\author{J. W. Negele\\
        Massachusetts Institute of Technology, Cambridge, Massachusetts 02139, U.S.A.\\
        E-mail: \email{negele@mit.edu}}
	
\author{A. Tsapalis\\
        Institute of Accelerating Systems and Applications, University of Athens, Athens, Greece\\
        E-mail: \email{a.tsapalis@iasa.gr}}			

\abstract{We develop a methodology that enables us to extract accurately the
          electromagnetic $\Delta$ form factors and their momentum dependence. 
          We test our approach in the quenched approximation as a preparation for
	  a study using dynamical fermions.
	  Our calculation of the four form factors
          covers pion masses between about 410 MeV and 560 MeV on lattices
          with a size of 2.9 fm and a lattice spacing $\sim\, 0.09$ fm. 
          From the form factors we are able to obtain estimates of 
          the magnetic moment and the charge radius of the $\Delta$, which 
          we compare to existing experimental and theoretical results.
          Our non-zero result for the electric quadrupole form factor 
          signals a deformation of the $\Delta$, pointing to an oblate
	  charge distribution.
          }

\FullConference{The XXV International Symposium on Lattice Field Theory\\
		 July 30-4 August 2007\\
		 Regensburg, Germany}

\begin{document}

\section{Introduction}
Electromagnetic form factors probe 
the structure of hadrons, yielding information on their size, shape and magnetization.
While the nucleon form factors and the $N\to\Delta$ transition
form factors have been studied quite thoroughly
both experimentally and on the lattice~\cite{Alexandrou:2006ru}, much less
has been done for the $\Delta$ form factors. Experiments are 
notoriously difficult due to the short mean life time of the
$\Delta$ of only about $6\times 10^{-24} s$. Nevertheless the magnetic 
moments of the $\Delta^+$~\cite{Kotulla:2002cg} and $\Delta^{++}$~\cite{LopezCastro:2000cv,Yao:2006px} have been 
measured. On the lattice the electromagnetic form factors of the $\Delta$ baryon have been evaluated 
at one value of the momentum transfer in a pioneer
quenched study in the early  nineties~\cite{Leinweber:1992hy}.
The main advantage of the fixed-sink approach that we adopt here, is that 
we are able to calculate the form factors for all values and directions
of the momentum transfer $\vec q$ simultaneously, allowing for an increased statistical precision.
Furthermore, we construct optimized sources that isolate the suppressed 
form factors enabling us, for the first time, to extract the electric 
quadrupole accurately.
Other improvements comprise reduced systematic 
errors due to finer lattice spacings, larger volumes and smaller pion masses.

\section{Lattice techniques}
We are working in Euclidean space-time throughout the whole paper.
We use Wilson fermions and the standard Wilson plaquette 
gauge-action. At all the pion masses considered in this work the $\Delta$
is a stable particle.

\subsection{Electromagnetic form factors of the $\Delta$ baryon}
The matrix element of the electromagnetic current, $V_\mu$, between two
$\Delta$-states can be decomposed in terms of four independent
covariant vertex function coefficients, $a_1(q^2)$, $a_2(q^2)$, $c_1(q^2)$ and $c_2(q^2)$,
which depend only on the momentum transfer squared $q^2=(p_f-p_i)^2$~\cite{Nozawa:1990gt}:
\begin{eqnarray}
  \langle \Delta(p_f,s_f) |\, V^\mu \, | \Delta(p_i,s_i) \rangle &=& \sqrt{\frac{m_\Delta^2}{E_{\Delta(\vec p_f)}E_{\Delta(\vec p_i)}}} \
                                                \bar u_\sigma(p_f,s_f)\, \Op^{\sigma \mu \tau}\, u_\tau(p_i,s_i) \\  
  \Op^{\sigma\mu\tau} &=& -\delta_{\sigma\tau}                \left[a_1 \gamma^\mu - i \frac{a_2}{2m_\Delta}P^\mu \right] 
                        +\frac{q^\sigma q^\tau}{4m_\Delta^2}\left[c_1 \gamma^\mu - i \frac{c_2}{2m_\Delta}P^\mu \right] \nonumber
   \, .
\end{eqnarray}
$E_\Delta$ and $m_\Delta$ denote the energy and the mass of the particle, $p_{i}\, (p_f)$ and $s_i\, (s_f)$ 
are the initial (final) four-momentum and spin-projection, while $P=p_f+p_i$. 
Every vector-component of the Rarita-Schwinger spinor $u_\sigma(p,s)$ satisfies the free
Dirac equation. Furthermore, the following auxiliary conditions are obeyed
\begin{equation}
   \gamma_\sigma u^\sigma(p,s) = 0, \qquad \qquad  
   p_\sigma u^\sigma(p,s) = 0 \, .
\end{equation}
The vertex function coefficients are linked to the phenomenologically more
interesting multipole form factors $G_{E0}$, $G_{E2}$, $G_{M1}$ and $G_{M3}$ by
a linear relation~\cite{Nozawa:1990gt}. The dominant form factors are the electric
charge, $G_{E0}$, and the magnetic dipole, $G_{M1}$, form factors.

\subsection{Interpolating fields}
We use an interpolating field that has the quantum numbers of the
$\Delta^+$ baryon
\begin{equation}\label{interpolatingfield}
   {\mathbf \chi}^{\Delta^+}_{\sigma\alpha}(x) = \frac{1}{\sqrt{3}} \epsilon^{abc}\Bigl[2\left({\mathbf u}^{a\top}(x) C\gamma_\sigma {\mathbf d}^b(x)\right) {\mathbf u}_\alpha^c(x)  
                                     + \left({\mathbf u}^{a\top}(x) C\gamma_\sigma {\mathbf u}^b(x)\right) {\mathbf d}_\alpha^c(x)\Bigr]\, ,
\end{equation}
where $C$ is the charge conjugation matrix.
To facilitate ground-state dominance we employ a covariant Gaussian smearing~\cite{Alexandrou:1992ti} on the
quark-fields entering Eq.~(\ref{interpolatingfield})
\begin{eqnarray}
        {\mathbf q }_\beta(t,\vec x) &=& \sum_{ \vec y} [\eins + \alpha H(\vec x,\vec y; U)]^n \ q_\beta(t,\vec y) \\
	H(\vec x, \vec y; U)  &=& \sum_{\mu=1}^3 \left(U_\mu(\vec x,t)\delta_{\vec x, \vec y - \hat \mu} + U^\dagger_\mu(\vec x-\hat \mu, t) \delta_{\vec x,\vec y+\hat\mu} \right)
\end{eqnarray}
Here $q$ is the local quark field (i.e. either $u$ or $d$), ${\bf q}$ is the smeared quark field and $U_\mu$ 
is the $SU(3)$-gauge field. 
For the lattice spacing and pion masses considered in this work, the values
$\alpha=4.0$ and $n=50$ ensure ground state dominance with the shortest
time evolution that could  be achieved.

\subsection{Correlation functions}
We specialize to a kinematical setup where the final $\Delta$-state is at rest ($\vec p_f =\vec 0$) and 
measure the two-point and three-point functions
\begin{eqnarray}
   G_{\sigma \tau}(\Gamma^\nu,\vec p, t_f-t_i) &=&\sum_{\vec x_f} e^{-i\vec x_f \cdot \vec p}\, 
   \Gamma^\nu_{\alpha'\alpha}\, \langle {\mathbf \chi}_{\sigma\alpha}(t_f,\vec x_f) \bar{\mathbf \chi}_{\tau\alpha'}(t_i, \vec 0) \rangle \label{twopoint} \\
   G_{\sigma\ \tau}^{\ \mu}(\Gamma^\nu,\vec q, t) &=& \sum_{\vec x,\, \vec x_f} e^{i\vec x \cdot \vec q}\,
   \Gamma^\nu_{\alpha'\alpha}\, \langle {\mathbf \chi}_{\sigma\alpha}(t_f,\vec x_f) V^\mu(t,\vec x) \bar{\mathbf \chi}_{\tau\alpha'}(t_i, \vec 0)\rangle \, , \label{threepoint}
\end{eqnarray}          
where $V_\mu$ is the symmetrized, conserved lattice electromagnetic current. We work 
with a representation of the Clifford-algebra in which $\gamma_4$ is diagonal. In this
representation our choices for the $\Gamma$-matrices are 
\begin{equation}
   \Gamma^k = \frac{1}{2}\left(\begin{array}{l l}\sigma^{(k)} & 0\\ 0 & 0\end{array}\right) \quad {\rm and}\quad \Gamma^4 = \frac{1}{2}\left(\begin{array}{l l}\eins & 0\\ 0 & 0\end{array}\right)\, ,
\end{equation}
with $k=1,\ldots , 3$ and $\sigma^{(k)}$ being the Pauli matrices.
The ratio
\begin{equation}
        R_{\sigma\ \tau}^{\ \mu}(\Gamma,\vec q,t) = \frac{G_{\sigma\ \tau}^{\ \mu}(\Gamma,\vec q,t)}{G_{k k}(\Gamma^4,\vec 0, t_f)}\ 
				         \sqrt{\frac{G_{kk}(\Gamma^4,\vec p_i, t_f-t)G_{kk}(\Gamma^4,\vec 0  ,t)G_{kk}(\Gamma^4,\vec 0,t_f)}
					            {G_{kk}(\Gamma^4,\vec 0, t_f-t)G_{kk}(\Gamma^4,\vec p_i,t)G_{kk}(\Gamma^4,\vec p_i,t_f)}}\, ,
\end{equation}
with implicit summations over the indices $k$ with $k=1,\ldots, 3$,
becomes time independent for large Euclidean time separations $t_f-t$ and $t-t_i$:
\begin{eqnarray}
   R_{\sigma\ \tau}^{\ \mu}(\Gamma,\vec q,t) \to \Pi_{\sigma\ \tau}^{\ \mu}(\Gamma,\vec q) &=& 
   \sqrt{\frac{3}{2}}\left[\frac{2 E_{\Delta(\vec q)}}{m_\Delta} 
                          +\frac{2 E^2_{\Delta(\vec q)}}{m^2_\Delta} 
                          +\frac{  E^3_{\Delta(\vec q)}}{m^3_\Delta} 
                          +\frac{  E^4_{\Delta(\vec q)}}{m^4_\Delta} \right]^{-\frac{1}{2}} \nonumber \\ 
   && {\rm tr}\left[\Gamma\, \Lambda_{\sigma\sigma'}(p_f) \Op^{{\sigma'}\mu{\tau'}} \Lambda_{\tau'\tau}(p_i) \right] \, .
\end{eqnarray}
The traces act in spinor-space and the Euclidean Schwinger-Rarita spin sum is given by
\begin{equation}
   \Lambda_{\sigma\tau}(p)  = - \frac{-i\slashed{p}+m_\Delta}{2m_\Delta}\left[
                               \delta_{\sigma\tau}-\frac{\gamma_\sigma\gamma_{\tau}}{3}
                               +\frac{2p_\sigma p_{\tau}}{3m_\Delta^2} 
			       - i \frac{p_\sigma\gamma_{\tau}-p_{\tau}\gamma_\sigma}{3m_\Delta} \right] \, .
\end{equation}

Since we are evaluating the correlator of Eq.~(\ref{threepoint}) using  
sequential inversions through the sink~\cite{Dolgov:2002zm}, 
a separate set of inversions is necessary for every choice of vector and Dirac-indices.
The total of $256$ combinations is beyond our computational resources, and hence we concentrate on a few 
carefully chosen combinations given below
\begin{eqnarray}
   \Pi_\mu^{(1)}(\vec q) &=& \sum \limits_{j,k,l=1}^3 \epsilon_{jkl}\Pi_{j\ k}^{\ \mu}(\Gamma^4, \vec q) \label{comb1} \\
   \Pi_\mu^{(2)}(\vec q) &=& \sum \limits_{k=1}^3 \Pi_{k\ k}^{\ \mu}(\Gamma^4, \vec q) \label{comb2} \\
   \Pi_\mu^{(3)}(\vec q) &=& \sum \limits_{j,k,l=1}^3 \epsilon_{jkl}\Pi_{j\ k}^{\ \mu}(\Gamma^j, \vec q) \label{comb3}\, .
\end{eqnarray}
From these all the multipole form factors can be optimally  extracted. For instance~(\ref{comb1}) 
is proportional to $G_{M1}$, while (\ref{comb3}) isolates $G_{E2}$ for $\mu=4$.

\subsection{Data analysis}
For a given value of $q^2$ the combinations given in Eqs.~(\ref{comb1}) to~(\ref{comb3}) are evaluated
for all different directions of $\vec q$ resulting in the same $q^2$, as well as for 
all four directions $\mu$ of the current. This leads to an over-constrained linear
system of equations, which is then solved in the least-squares sense yielding 
estimates of $G_{E0}$, $G_{E2}$, $G_{M1}$ and $G_{M3}$. This estimation is
embedded into a jackknife binning procedure, 
thus providing  statistical 
errors for the form factors that take all correlation and 
autocorrelation effects
into account.

\section{Results}
\subsection{Simulation parameters}
The calculation has been performed on  a lattice of volume $L^3 \times T = 32^3 \times 64 $ 
in the quenched approximation. The lattice spacing has been estimated
using the nucleon mass in the chiral limit~\cite{Alexandrou:2006ru} to 
be~$a=0.092(3)$, so 
the spatial extent of the lattice is~$\sim\, 2.9$~fm. 
We work with two degenerate Wilson valence quarks. 
Isospin symmetry relates  results obtained for the $\Delta^+$ to 
those for the $\Delta^{++},\, \Delta^0$ and $\Delta^-$ since they differ only
by a charge-factor.
The pion and $\Delta$ masses corresponding
to the three values of the hopping parameter, $\kappa$, considered here are summarized in Table~\ref{resultstable}.
We use 200 well separated gauge configurations for the calculation.
Contributions from disconnected diagrams are neglected.
The numbers and figures below are obtained from the connected contributions to the
full electromagnetic current. They also can be interpreted as the iso-vector
form factors with the iso-vector current $V_\mu^{I} = \bar u \gamma_\mu u - \bar d \gamma_\mu d$.  

\subsection{Electric charge form factor $G_{E0}$}
Our results for $G_{E0}$ are displayed in the left panel of
Fig.~\ref{ge0gm1fig}. The momentum dependence of our data is 
described very well by the dipole ansatz
\begin{equation}\label{ge0dipoleform}
   G_{E0}(q^2) = \frac{1}{(1+ c_{E0}\, q^2)^2} \, .
\end{equation}
Non-relativistically the slope at $q^2=0$ is related to the electric 
charge radius via
\begin{equation}\label{chargeradius}
   \left\langle r^2 \right\rangle = -6 \left. \frac{d}{dq^2} G_{E0}(q^2)\right|_{q^2=0} \, , 
\end{equation}
for which the results are collected in Table~\ref{resultstable}.
An extrapolation of the form factor to the physical pion mass, linear in $m_\pi^2$,
is used to obtain the last row of the table.
One could consider using chiral perturbation theory for the extrapolations~\cite{Arndt:2003we}.
This will be done when we have results using dynamical fermions.

After a chiral extrapolation, linear in $m_\pi^2$,  the authors of ref.~\cite{Leinweber:1992hy} obtained 0.63(7) fm, which
is compatible with our value of 0.691(6) fm.

\subsection{Magnetic dipole form factor $G_{M1}$}
\begin{figure}[H]
   \hspace{-0.5cm}\includegraphics[width=0.53\linewidth]{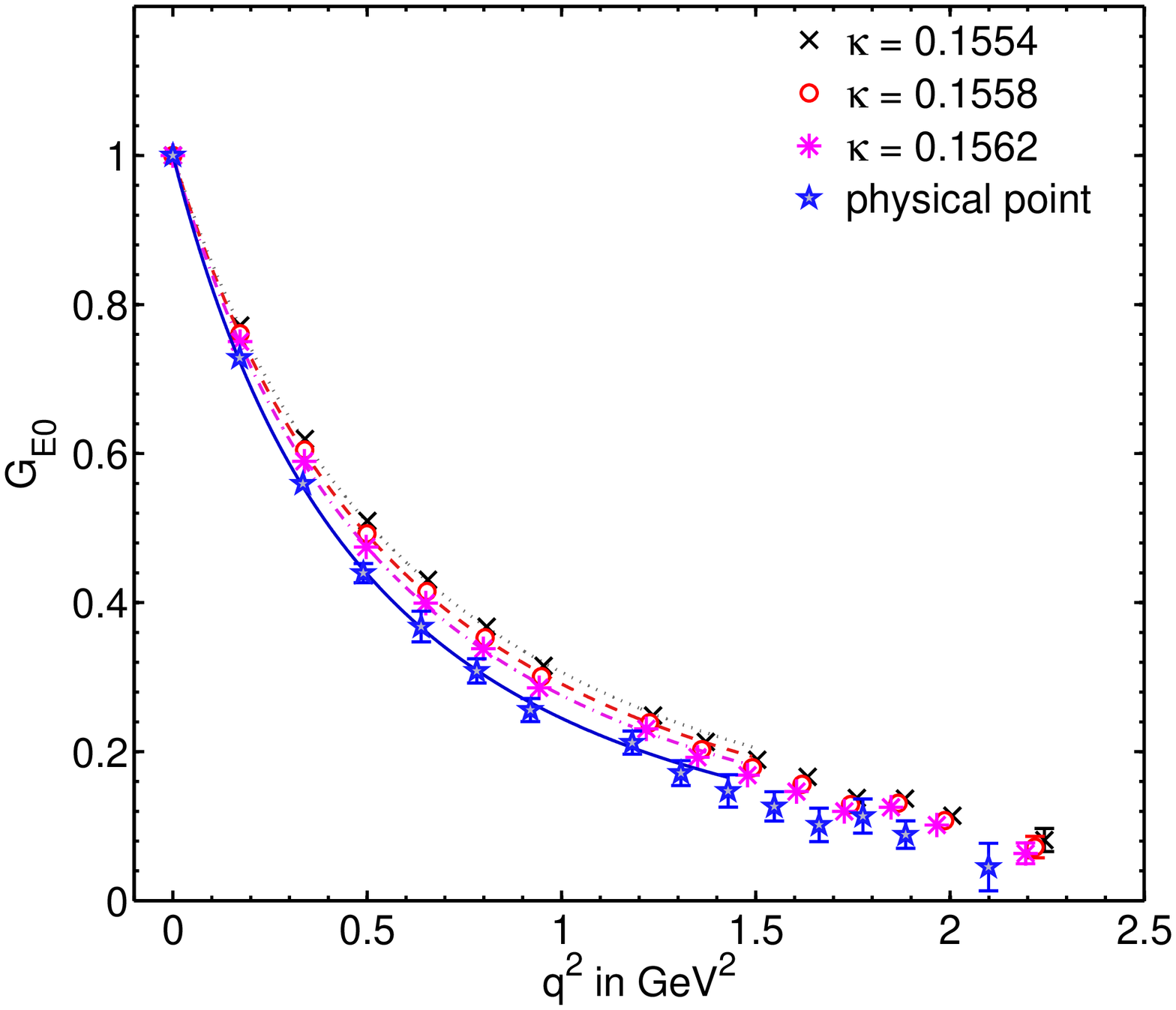}
   \hspace{-0.2cm}\includegraphics[width=0.53\linewidth]{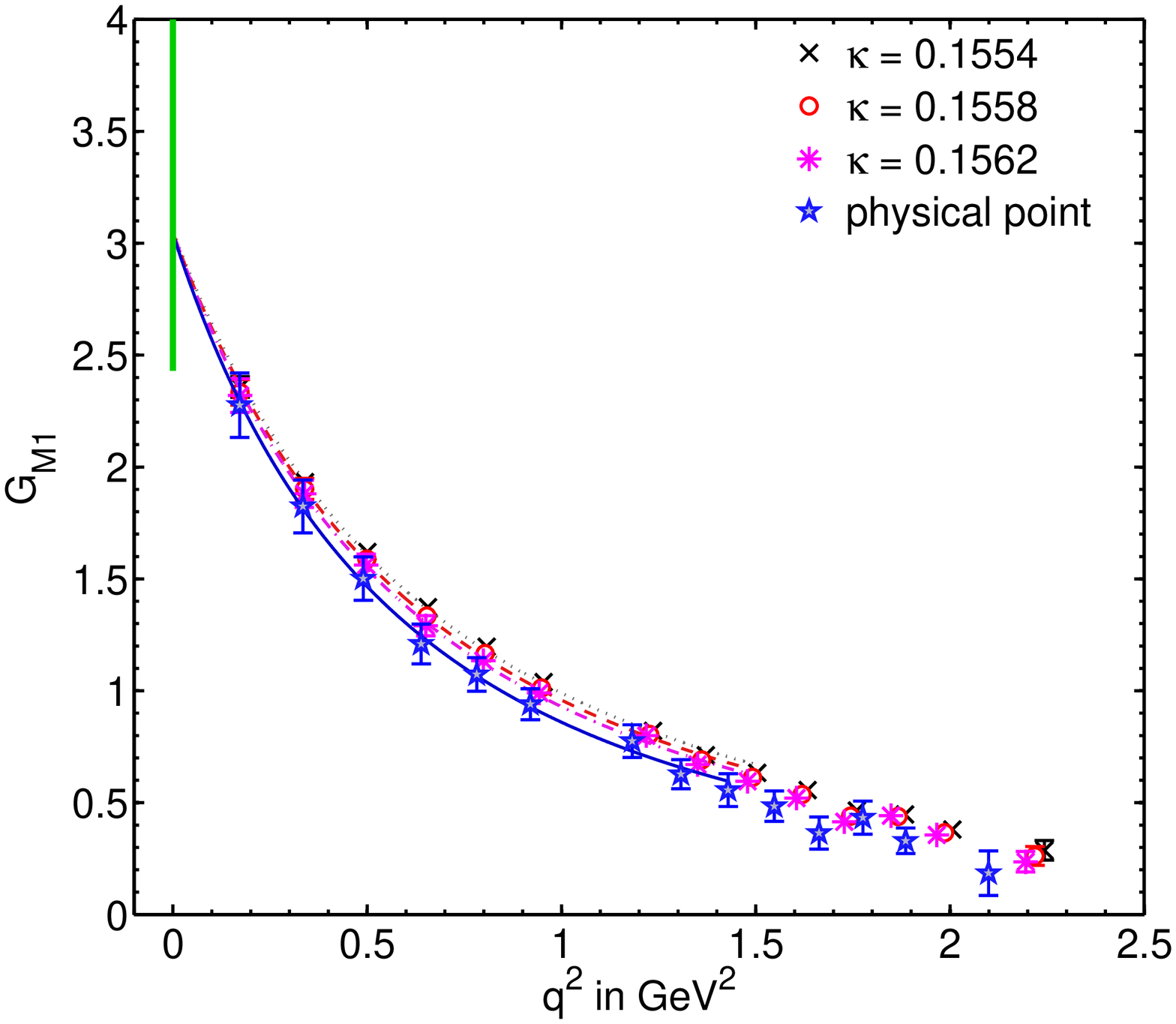}
   \caption{The two dominant form factors for the three pion masses as well as
            the result of a linear chiral extrapolation to the physical pion mass. 
            The left panel shows the electric charge form factor, $G_{E0}$, and
	    the right panel shows the magnetic dipole form factor, $G_{M1}$.
	    The lines are fits to the dipole forms Eqs.~(3.1) and~(3.3).
	    The band at $q^2=0$ indicates the experimentally measured value as 
            quoted by the particle data group.
	    }\label{ge0gm1fig}
\end{figure}

Also $G_{M1}(q^2)$ can be fitted to a dipole ansatz
\begin{equation}\label{gm1dipoleform}
   G_{M1}(q^2) = \frac{a_{M1}(0)}{(1+c_{M1}\, q^2)^2} \, .
\end{equation}
From the fit parameter $a_{M1}$
the magnetic moment $\mu_{\Delta^+}= G_{M1}(0) \left(\frac{e}{2m_\Delta}\right)$ 
of the $\Delta$ can be determined. It is given in Table~\ref{resultstable}. 
Fits and data are shown in Fig.~\ref{ge0gm1fig}. 
\vspace{-0.3cm}
The magnetic moment of the $\Delta^{++}$ has been measured in several experiments. 
However not all experimental values are in agreement with each other,
so the particle data group quotes a rather
broad band of $3.7 \mu_N$ to $7.5\mu_N$ ~\cite{Yao:2006px}. The most 
recent experiment~\cite{LopezCastro:2000cv}
quotes $\mu_{\Delta^{++}}=6.14(51) \mu_N$. Our estimate of 
$\mu_{\Delta^{++}}=2 \mu_{\Delta^+} = 4.64(32) \mu_N $ lies in the middle of 
the error band given by PDG
and below the result of~\cite{LopezCastro:2000cv}.
An experiment~\cite{Kotulla:2002cg} that measures the magnetic
moment of the $\Delta^+$ arrives at $\mu_{\Delta^+} = (2.7 \pm 1 \pm 1.5 \pm 3) \mu_N$\, . 
Also recent lattice calculations yield results on the 
magnetic dipole moment. 
The value of $1.6(3)\mu_N$ for the $\Delta^+$, read off from fig.~6 of Ref.~\cite{Boinepalli:2006ng}
at a pion mass of around 300 MeV lies slightly below our value of $2.32(16) \mu_N$.

\subsection{Electric quadrupole form factor $G_{E2}$}
This is the first accurate evaluation of the quadrupole form factor, $G_{E2}$,
within lattice QCD. 
The intrinsic quadrupole moment $Q = m_\Delta^{-2} G_{E2}(0)$,
non-relativistically, is given by~\cite{Leinweber:1992hy}
\begin{equation}
    Q = \int \! d^3r \ \bar \psi(r) \left( 3z^2-r^2\right)\psi(r) \, , 
\end{equation}
where $\psi$ is the wave function of the $\Delta$.
A negative value, as that shown in Fig.~\ref{ge2gm3fig}, corresponds to 
an oblate deformation of the $\Delta$.

The value $G_{E2}(0) = -0.4(14)$ obtained in reference~\cite{Leinweber:1992hy} is compatible with ours,
but comes with a much larger statistical error.

\subsection{Magnetic octupole form factor $G_{M3}$}
\begin{figure}[H]
   \hspace{-0.5cm}\includegraphics[width=0.53\linewidth]{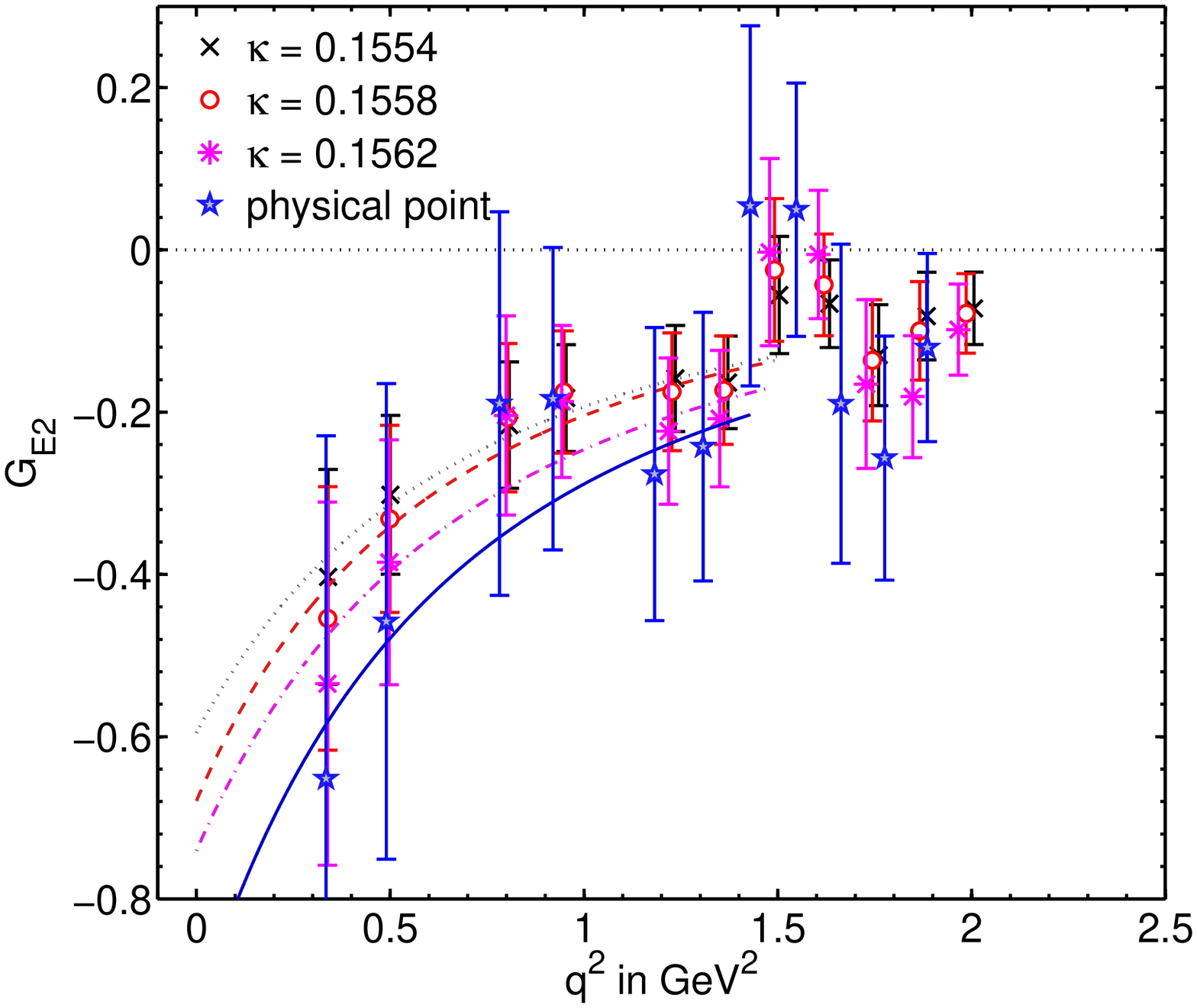}
   \hspace{-0.2cm}\includegraphics[width=0.53\linewidth]{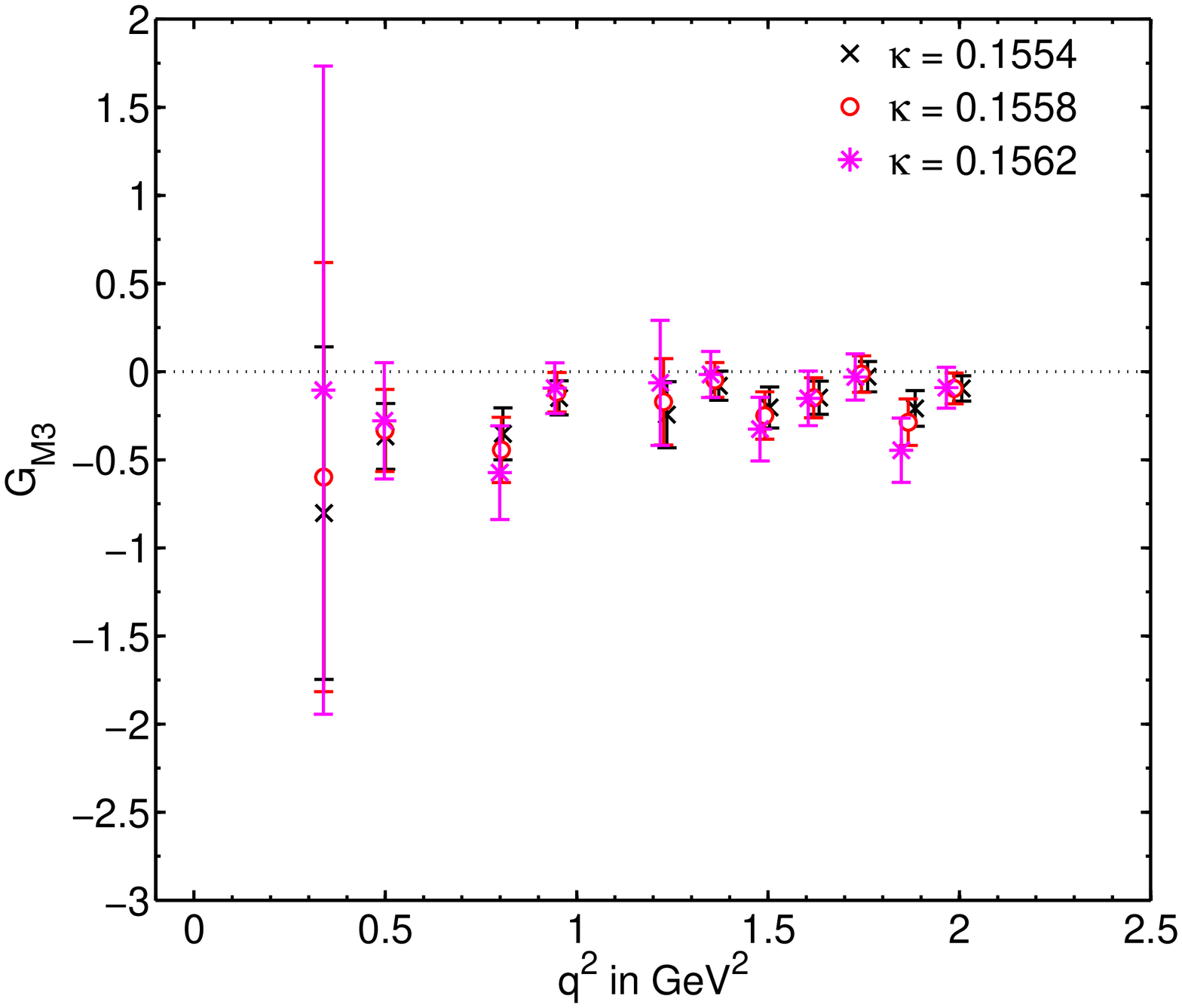}
   \caption{The electric quadrupole form factor (left panel) and the
            magnetic octupole form factor (right panel). The notation is
	    the same as in~Fig.1.
           }\label{ge2gm3fig}
\end{figure}

Our signal for $G_{M3}$ is not very strong, as can be seen in Fig.~\ref{ge2gm3fig}. 
This form factor has a small 
 negative value, albeit with large  statistical errors that,  
at present, do not exclude a zero value. To next to leading order of a chiral 
expansion a value of zero is expected~\cite{Arndt:2003we}.

\vspace{-0.3cm}

\begin{table}[h]
  \centering
  \begin{tabular}{l l l l l l}
  \toprule
  $ \kappa   $&$ m_{\pi}\ [MeV] $&$ m_\Delta  $&$ \langle r^2 \rangle^{\frac{1}{2}} $&$ G_{M1}(0) $&$ G_{E2}(0) $\\
  \midrule
  $ 0.1554   $&$ 563(4)         $&$ 1.470(15) $&$  0.614(2)                         $&$ 3.05(7)   $&$ -0.6(3) $\\
  $ 0.1558   $&$ 490(4)         $&$ 1.425(16) $&$  0.632(2)                         $&$ 3.05(8)   $&$ -0.7(4) $\\
  $ 0.1562   $&$ 411(4)         $&$ 1.382(15) $&$  0.650(3)                         $&$ 3.05(10)  $&$ -0.7(4) $\\
  \midrule
  $          $&$ 135            $&$           $&$  0.691(6)  $&$ 3.04(21)$&$  $\\
  \bottomrule
  \end{tabular}
  \caption{Estimated values of the $\Delta$ mass,  the charge radius, $G_{M1}(0)$ and $G_{E2}(0)$ for the
           pion masses considered. The last row contains our results from data extrapolated
           to the physical value of the pion mass.}\label{resultstable}
\end{table}

\section{Conclusions and outlook}
Our results on the electromagnetic form factors of the $\Delta$
baryon and their momentum  dependence confirm the up to
now phenomenological description, e.g. the $q^2$-dependence of
$G_{E0}$ and $G_{M1}$ are well described by a dipole ansatz.
A particularly interesting result of our study is 
the non-vanishing, negative value of the electric quadrupole
form factor, which is associated with an oblate $\Delta$.

While the statistical errors in our calculation are under control, there
are several sources of systematic errors that we would like to address.
The results presented here were obtained in the quenched approximation. 
A study with dynamical fermions is currently in progress. 
It is not clear how strongly the neglect of disconnected contributions 
affects our final results. This is a problem we share with most other
calculations of electromagnetic form factors. Only recently, with new methods and faster
machines, we are beginning to address the calculation of the disconnected contributions
in a statistically controlled manner.

While the finite volume effects are expected to be negligible on our rather big
lattice, cutoff effects may be significant,  especially towards higher
momentum transfers.

\end{document}